# Abundance-Distributions in Artificial Life and Stochastic Models: "Age and Area" revisited

Chris Adami, C. Titus Brown and Michael R. Haggerty

W.K. Kellogg Radiation Laboratory
California Institute of Technology
Pasadena, CA 91125

**Abstract.** Using an artificial system of self-replicating strings, we show a correlation between the age of a genotype and its abundance that reflects a punctuated rather than gradual picture of evolution, as suggested long ago by Willis. In support of this correlation, we measure genotype abundance distributions and find universal coefficients. Finally, we propose a simple stochastic model which describes the dynamics of equilibrium periods and which correctly predicts most of the observed distributions.

## 1 Introduction

Species-abundance distributions have played an important role in our understanding of the process of evolution on the one hand, and in the field of ecology on the other. Early on, Willis [1] remarked that the frequency distribution of species within genera is markedly concave, i.e. there are many genera with very few species but only a few with very many species. In fact, it was Willis' objective to disprove Darwin's claim that new species arise through the "survival of the fittest," in a scenario where whole genera adapt gradually. In such a scenario, there is no correlation between the age of a genus and the number of extant species. A genus with many species may be old, or it may have adapted as a whole, gradually, until the differences are so large that a new genus is formed. Then, even though the genus would be considerd young [2], it would show a lot of variation. Willis, on the contrary, believed that mutation acts on the individual, and that any "young" genus will have, on average, less speciation then an "old" one.

This view, of the origin of species *per saltum*, i.e. the creation of new genera by extremely rare mutations that trigger an avalanche of speciation, as opposed to the gradual adaptation of species through natural selection, was taken up by Yule [3] in a remarkable paper. He developed a mathematical theory of evolution based on this picture, which matched the species-abundance curves obtained by Willis with high accuracy. The theory proposed was simple. For one ancestral genus, he assumes a certain probability for a "specific" mutation which creates a new species, as well as a (smaller) probability for a mutation that gives rise to a new genus: a "generic" mutation. Iterating this process, he predicts (in the limit of large number of species and infinite evolutionary time) a distribution for the number of genera $N_g$ with $n$ species

$$N_g(n) \sim \frac{1}{n^{1+1/\rho}} \;, \tag{1.1}$$

where $\rho \geq 1$ is the ratio between the probability for a "specific" to a "generic" mutation. This allowed him to fit most observed distributions with a parameter $\rho \approx 2$, which fit the species-abundance relations available to him (see Fig. 1).

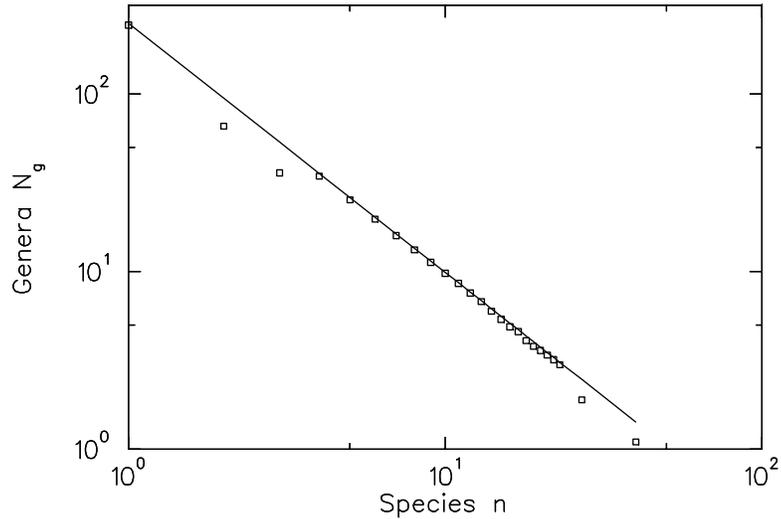

Fig. 1. Number $N_g$ of genera (binned) of *Leguminosae* with $n$ species, compiled by Willis from "Dictionary of Flowering plants," as quoted by Yule [3] (errors are unavailable). The solid line is Yule's model-fit (infinite time) for $\rho = 2.457$, i.e. $D_s = 1.4$. The "finite time" fit raises the power to $D_s = 1.5$.

Furthermore, he concluded from his model that there is indeed, as postulated by Willis, a relationship between age and "size" of a genus, and therefore that evolution proceeds by leaps and bounds, rather than by gradual Darwinism. Seventy years later we have much sympathy with this interpretation, since punctuated equilibrium as promoted by Gould and Eldredge (see, for example, [4]) has become the standard evolutionary picture.

Ecologists, on the other hand, are interested in the abundance-distribution of species in a specified area, to understand the mechanisms that govern influx of new species and the extinction of others. The connection between these apparently unrelated distributions is established via the age/"size" correlation of species or genera. Should such a correlation exist, the ecological distribution reflects the dynamics of competition rather than a distribution of resources, and those species taking up more space in the area studied are simply the older ones. In the following, we would like to argue that, not only is there such a relationship in the simple Artificial Life system that we investigate, but that most species-abundance distributions arise from very simple underlying dynamics governed by random processes. We suggest that "taxon"-abundance curves, independent of the placement in the hierarchy of taxonomic groups, should show universal power laws with coefficient (-3/2), based on the model presented in section 3. In fact, there is some evidence supporting this point of view from the work of Burlando [14], who compiled abundance-distributions from taxonomic data and the fossil record, and observed power-laws with coefficients consistent with the one obtained here.

Previously [8, 9], it has been suggested that such power-laws can be understood if living systems are in a self-organized critical state. In this paper, we proceed from a different point of view. We measure, in an Artificial Life system, the most basic of the abundance distributions, that of genotypes. We try to establish that the power-law obtained there is roughly universal, by studying its dependence on system size and mutation rates. We establish that the power-law is observed for the "instantaneous" (i.e. ecological) distribution, by taking snapshots of it in time, and the "integrated" (i.e. historical) distribution, which determines the "size" of a specific genotype. We use the latter to establish the "Age-Area" relation. In the next step, we present a simple model that reproduces most of the features observed in genotype-abundance distributions, but which in principle can explain the gross features of abundance distributions of *any* taxon, instantaneous and integrated, simply by relabelling the probabilities.

## 2 Abundance-Distributions and Artificial Life

Several models for the species-abundance curve have been proposed, and fitted to data obtained from a variety of fauna and flora. Some, like MacArthur's "broken-stick" model [11], relate the species-abundance distribution to the distribution of resources and niches, while others invoke simple branching and mutation models (without competition) such as Yule's, predicting geometric series (and power laws in the limit of a large number of species). Others include the effects of competition (such as [12] and much later [13]), and obtain frequency distributions ranging from exponential to power-law.[1]

Here, we take a novel approach to the determination of species-abundance relations, made possible by the advent of pioneering Artificial Life systems such as the tierra system, developed by Ray [5], and the avida system [10], developed by our group at Caltech. Keeping in mind the caveats mentioned in the Introduction, we investigate genotype frequency distributions from a system of self-replicating bit-strings subject to mutation and survival of the fittest. Each codon (from an alphabet of 20) codes for an instruction in a special machine-language for simple "programs" running on virtual CPUs. The language permits programs that self-replicate, and the chemistry of self-replication is thus substituted by the execution of the program (see [5, 7, 10] for more details on the tierra and avida systems). The aspects of the statistical mechanics of self-replication of the bit-strings have already been investigated with one approach [9], and will be investigated with another in this paper; such theoretical understanding allows us to make predictions and test them against the experimental results obtained with the Artificial Life systems. How much this artificial system resembles the global behavior of populations of self-replicating RNA will become known once such natural systems become available.

### 2.1 Dynamics of Self-Replicating Strings

Any string in the population is characterized by its specific sequence of instructions, which is termed the *genotype* of the string. If a genotype replicates accurately, we can associate a *replication rate* $\epsilon$ (number of offspring per unit time) with it, and the string then competes with neighboring strings for the placement of offspring. In avida, strings (or "cells") are arranged on a two-dimensional grid of fixed size and the total number of cells is constant throughout the run. When a new cell is spawned, the offspring replaces the oldest of the nine cells in the neighborhood of its parent. This mechanism of placing offspring in nearest-neighbor sites (thereby removing potentially competing cells) constitutes the only significant method of interaction between cells in this system, and results in the dissipative transport of information contained in the genome throughout the population (see [10] for details on this system).

---

[1] see [13] for a brief review of abundance-rank relations.

Strings are subjected to Poisson-random "cosmic ray" mutations that replace instructions randomly at an average rate $R$ (mutations per site per unit time) such that the probability for a string of length $\ell$ to be hit by a mutation is $R\ell$. In the first approximation, then, genotypes are governed by the following "kinetic equation," which models the non-stochastic aspects of the time development of the occupation number (or frequency) $n_i(t)$ of genotype $i$:

$$n_i(t+1) - n_i(t) = (\epsilon_i - \langle\epsilon\rangle - R\ell)n_i(t) + C, \qquad (2.2)$$

where $\epsilon_i n_i$ is the average number of cells of genotype $i$ born per unit time, and $\langle\epsilon\rangle n_i$ is the average number that die (when cells of different genotypes replicate into their spot). The flux term $C$ models cells mutated into genotype $i$ from "mutationally close" genotypes $j$, and can be neglected in most situations.

In that case, and when the average replication rate $\langle\epsilon\rangle$ is roughly constant, Eq. (2.2) describes exponential growth or decline

$$n_i(t) = n_i(0)e^{\gamma_i t} \qquad (2.3)$$

where $\gamma_i$ is the growth factor $\epsilon_i - \langle\epsilon\rangle - R\ell$ of genotype $i$. Any genotype with $\gamma < 0$ experiences exponential decline and is soon pushed into extinction. A newly created genotype with a replication rate better than the old average, on the other hand, will experience exponential growth, quickly taking over the soup and temporarily reducing the diversity of the population until mutations restore it (see [8, 9] for a more complete description of the dynamics). In practice, we find that the system spends most of its time not in either exponential regime, but rather in equilibria where most genotypes have $\gamma_i \approx 0$. It is this third regime which turns out to dictate most of the properties of genotype distributions; the properties of this regime will be understood below through the use of the DL model.

## 2.2 Results

The empirical data which will be presented below were produced in a number of avida simulations. We have measured the number of genotypes $n_i(t)$ for each living genotype at different times (every ten updates), effectively taking "snapshots" of the genotype-abundance distribution, and averaged the 3,000 snapshots for each of twenty runs for populations of sizes 20x20 and 40x40, as well as 3,000 snapshots for each of five runs of size 80x80, all at an intermediate mutation rate $R = 40 \times 10^{-5}$ mutations per genome site per update.

The distributions $N_g(n)$ of genotypes with $n$ living copies are shown in Fig. 2a as log-log plots. The slope of a straight line in a log-log plot determines the exponent in a power law, and we obtain from these measurements distribution functions of the form

$$N_g(n) \sim \frac{1}{n^{D_s}} \qquad (2.4)$$

with $D_s$ between 1.45 and 1.85 for the different sizes.

The peculiar rise in the distribution close to the maximum population size is due to the finite size of the lattice. The genotypes accumulating there are in fact the few ones that enjoy exponential growth after an invention that gives them an edge over all extant genotypes.

We are also interested in the dependence of the distribution function on the mutation rate. For the 20x20 system, we have measured the distribution function at half and twice the mutation rate used in Fig. 2a; the results can be seen in Fig. 2b. The power-law exponents $D_s$ for the three mutation rates fall between 1.45 and 1.64.

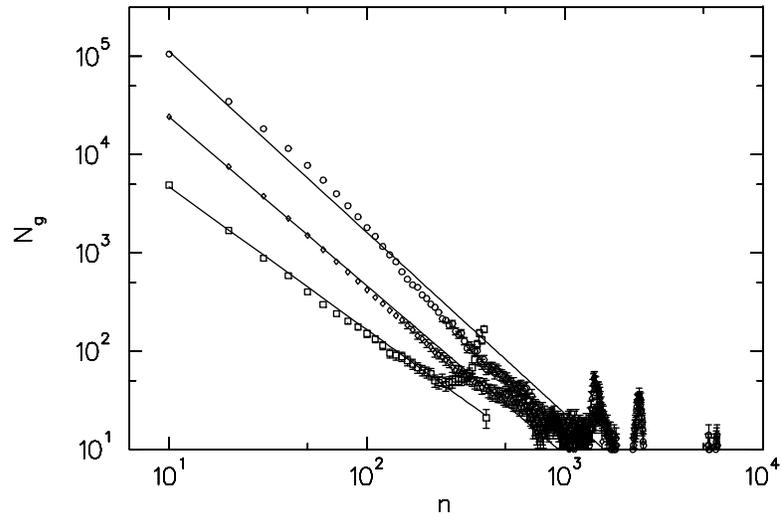

Fig. 2a. Genotype-abundance distributions (number of genotypes with $n$ living copies), for three different population sizes in avida: 20x20 (squares), 40x40 (diamonds), and 80x80 (circles). The straight lines are least-squares fit to the data with power-law coefficients of 1.45, 1.7, and 1.84 respectively.

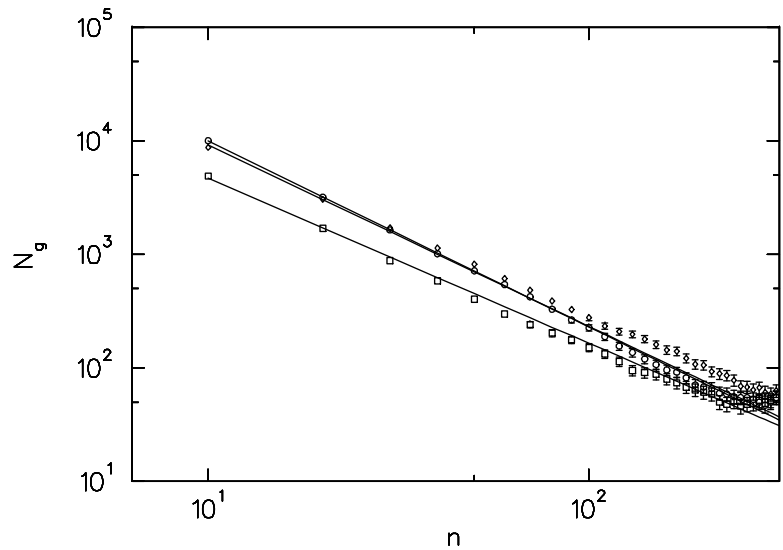

Fig. 2b. Genotype-abundance distributions (number of genotypes with $n$ living copies), for three different mutation rates in avida (small: squares, medium: diamonds, high: circles) and a population size 20x20. The straight lines are least-squares fit to the data with power-law coefficients 1.45, 1.60, and 1.64 respectively.

At the same time we measured the distribution of *genotype ages*, by recording, for each genotype, the time between creation (via mutation) and extinction. It was Willis' idea that there ought to be a correlation between genotype ages and sizes, if new species are created (as we now know) by an extremely rare mutation of *one* copy of an existing one. We can check this hypothesis by plotting genotype sizes (total number of cells of a genotype that ever lived, analogous to the total number of species that were ever produced by a genus) versus the age of that genotype (Fig. 3). As expected, we see a correlation between genotype size and age (the "Age-Area" correlation), where most of the points that lie close to the diagonal are in fact genotypes with a vanishing growth factor. A few genotypes in each run start out with a positive growth factor, i.e. a large fitness gradient. These are precisely those genotypes that form the nucleus of a new "generation" or "species," and therefore grow much faster than average genotypes. For the run used to produce Fig. 3, they have been marked by diamonds, and they show a large deviation from the rest of the Age-Area curve.

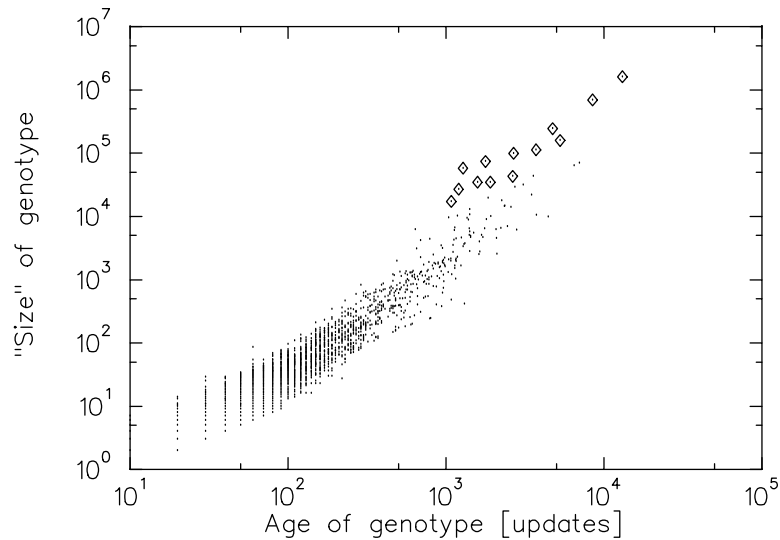

Fig. 3. Genotype "size," as measured by sampling *integrated* frequency over time, as a function of genotype age (in units of population updates). The genotypes marked by diamonds had sizable growth factors, and represent mutations that ushered in a new epoch. As such, they deviate from the average genotypes which have zero growth factors and are described by the DL model (below).

In the next section, we try to understand these results by comparing them to the simplest model that can be used to describe species or genotype distributions.

## 3  The DL Model

Analysis of genotype frequency curves as a function of time as obtained from the avida system reveals that these curves share many characteristics with a simple random walk. While of course this cannot be the whole story (for example, it lacks the dynamics of interaction between genotypes present in avida), it turns out nonetheless that this simple "dumb luck" model of equilibrium genotype dynamics reproduces much of the universal behavior seen in the full avida model.

In the DL model, a species is born with a single member, when an individual of another species suffers a mutation. The new species is characterized by a rate of growth, $R_g$, and a rate of shrinkage, $R_s$, which are assumed to be constant. We rule out unbounded exponential growth by making the additional assumption that $R_g \leq R_s$.

From that point on, the population $n(t)$ of the species increases or decreases with randomly occurring single births or deaths, given for a small time $\Delta t$ by the probabilities

$$P_g \equiv P(\text{one birth}) = R_g n(t) \Delta t, \tag{3.5}$$

$$P_s \equiv P(\text{one death}) = R_s n(t) \Delta t. \tag{3.6}$$

Both births and deaths are necessarily assumed to be proportional to the number of individuals alive at that time (as in (2.2)). The first time that $n(t)$ becomes zero again, the species is extinct. The key point is that the number of individuals is always an integer, and changes in discrete steps according to the Poisson process described. The population $n(t)$ thus takes a random walk—prosperity or plague determined by dumb luck, modified only by an overall species fitness parameter $(R_g/R_s)$. As it turns out, this model seems to describe most of the equilibrium population dynamics of species in avida, where creature interactions are relatively unimportant and mixing takes place on a shorter time scale than evolution.

Trivially, the long-time behavior of a species in this model is exponential decline whenever $R_g < R_s$, and that behavior ensures that every species will eventually become extinct. However, we are interested in the dynamics in the regime of low population counts, where the stochasticity of the model plays the dominant role. Therefore we have written a simulation of genotype dynamics following the DL model to compare with avida data. Some typical population time series of this simulation are shown in Fig. 4.

### 3.1 Comparison with a positive random walk

The classical random walk problem is similar to the DL model, and studying it will give us some insight into the behavior of our system. In a random walk, there are constant probabilities $P_g$ and $P_s$ (with $P_g + P_s = 1$) of taking unit steps in the positive or negative directions, respectively, at each unit of time. If we choose variable time steps of

$$\Delta t = \frac{1}{n(t)(R_g + R_s)}, \tag{3.7}$$

in the equations above, then the proportionality of growth and shrinkage rates to $n$ is attained. If we further constrain the random walk to positive values (extinction being identified with the first time the walk returns to zero), the models are alike in their essential characteristics.

For the positive random walk, the probability of ending up at position $n$ after exactly $N$ steps can be calculated analytically; it is

$$P(N, n) = \begin{cases} \frac{2n}{P_g N} \binom{N}{(N+n)/2} P_g^{(N+n)/2} P_s^{(N-n)/2} & \text{for } n > 0 \\ \frac{1}{2(N-1)P_g} \binom{N}{N/2} P_g^{N/2} P_s^{N/2} & \text{for } n = 0, \end{cases} \tag{3.8}$$

where $N + n$ must be even. Although the expression does not hold rigorously for the DL model, it is a good approximation and will help illuminate several points.

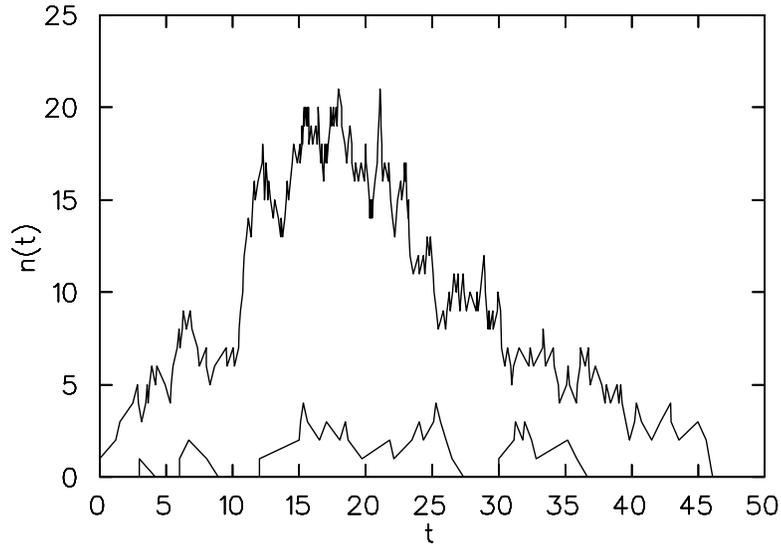

Fig. 4. Time-series of 5 independently run DL simulations of genotype abundances. Each species has the same fitness $R_g/R_s = 0.999$, so the surprising success of the one can only be attributed to "dumb luck."

## 3.2 Total historical genotype populations

The positive random walk and DL models make identical predictions for the distribution of the total historical genotype population (the total number of individuals of a genotype that have ever been born). Since each time step coincides with either a single birth or a single death, the sum of births and deaths equals the number of time steps. Since the genotype starts and ends with zero members, moreover, the number of births and the number of deaths must be equal. The probability that a genotype has a total number $N_b$ births during its whole history is therefore given by $P(2N_b, 0)$ from Eq. (3.8)—for large $N_b$, approximately

$$P(2N_b, 0) \approx \frac{1}{P_g 4\sqrt{\pi}} \frac{1}{N_b^{3/2}} (4P_g P_s)^{N_b}. \qquad (3.9)$$

This expression has two interesting limits. First, for creatures with low fitness ($P_g \ll P_s \approx 1$), there is an exponential suppression $(4P_g)^{N_b}$ of large historical populations (see Fig. 5). This is nothing more than the tendency of an unfit species towards exponential decline. This is one reason why unfit creatures do not have a big effect in observed data.

In a realistic avida run, however, mutations that are not fatal are usually neutral, and thus the most important case is $P_g \approx P_s \approx \frac{1}{2}$. In this limit, the exponential factor is negligible, and the distribution is dominated by the power law $N_b^{-3/2}$. This is indeed the behavior seen in the DL model simulations (see Fig. 5), avida population data (Figs. 2a,b), and also in many studies of biological species abundances [3].

The universal exponent ($-3/2$) relies only on the assumption that populations of species are noninteracting; it is a property of any species which evolves stochastically. In particular, it does not depend on the mutation rate (rate of creation of new genotypes), in contrast to the model of Yule. Thus, the fact that his fits to observational data produce an exponent of roughly ($-3/2$) should not be seen as a property of the biological systems that he studied, but rather as a universal property accidentally uncovered in his analysis of the data.

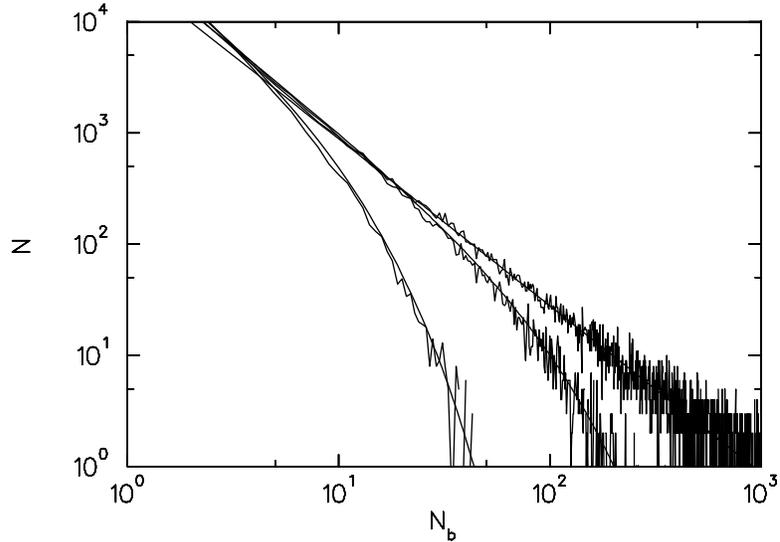

Fig. 5. Distribution of total number $N$ of random DL walkers with $N_b$ total births, with fitness parameter $\rho = R_g/R_s = 0.999$ (upper curve), 0.8 (middle curve), and 0.5 (lower curve). The distributions are fit using the parameters obtained by the theoretical estimate $N \propto N_b^{-3/2}[4\rho/(1+\rho)^2]^{N_b}$ above. This models the "integrated," or "historical," distributions.

### 3.3 Instantaneous abundance distribution

The distribution of instantaneous abundances $n(t)$ for the positive random walk is found to be roughly exponential: $P(n) \propto \exp(-\alpha n)$. In this case the connection to the DL model can be made exactly: since its time steps are dilated by a factor of $n(t)$ during periods of higher abundances, those points are systematically oversampled by the same factor, and the instantaneous abundance distribution for the DL model is roughly

$$P(n) \propto \frac{e^{-\alpha n}}{n} \qquad (3.10)$$

(see Fig. 6a).

This distribution is heavily influenced by one factor present in avida but not included in the naive DL model: the effect of genotypes with high fitness ($P_g > P_s$), which appear occasionally and take over a large fraction of the soup. While neutral genotypes spend exponentially little time at high abundances, high fitness genotypes spend much of their time there. Therefore, when comparing avida data to theory, we omitted in each case the contribution of any genotype which ever filled more than 10% of the soup; this removes most of the high-fitness genotypes. (The cut also removes a significant number of neutral growth rate genotypes that happen to cross that threshold, but it does so equally for both data sets.) The result is that theory matches avida data quite well, as seen in Fig. 6b.

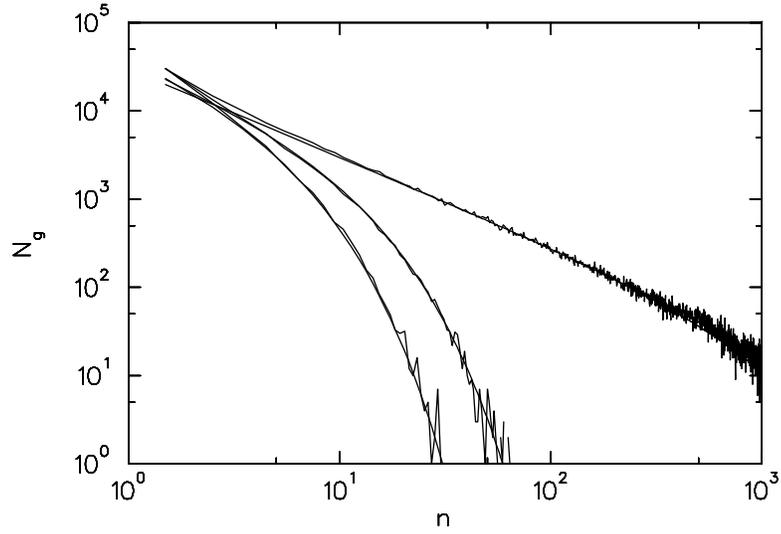

Fig. 6a. Genotype abundance distribution in a DL model obtained by sampling genotype frequencies (such as in Fig. 4) every 10 time units ("instantaneous," or "ecological" distribution). The upper curve was obtained for a simulation of genotypes with $\rho = P_g/P_s = 0.999$ (almost neutral growth rates); the middle curve, $\rho = 0.9$; and the lowest curve, $\rho = 0.8$ (significantly inferior). The solid lines are fits to the data, of the form $N_g(t) \sim \exp{-\alpha n}/n$, with $\alpha$ parameters 0.001, 0.11, and 0.24 respectively.

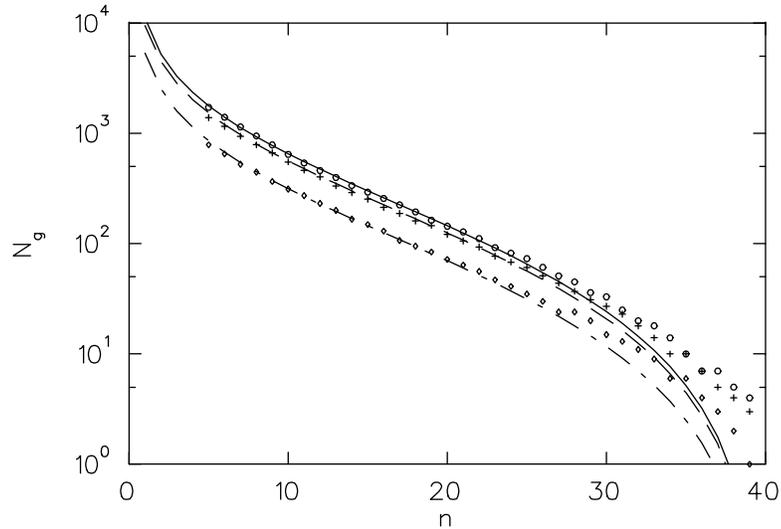

Fig. 6b. Genotype abundance distribution (ecological) from avida, omitting the "leading," fast replicating (high-fitness) creatures and compensating for that skewing on the neutral fitness abundance distribution. We have plotted here the genotype abundance distribution for genotypes with less than 40 members, in a 40 × 40 population, for small (diamonds), medium (crosses), and large (circles) mutation rates, semi-logarithmically. The short-dashed, long-dashed, and solid lines are the theoretical estimates from the DL model, with $R_s = R_g$ and no other free parameters except the normalization.

### 3.4 Age-area in the DL model

One can also make an age-area plot, comparable to Fig. 3, with data from the DL model simulation (see Fig. 7). Again, the data show that larger genotypes are older ones, consistent with evolution by infrequent jumps as opposed to gradual evolution. Comparison with Fig. 3 also highlights the fact that the vast majority of genotypes are neutrally fit, and explained well by the DL model. The few that lie above the "neutral" curve are the exceptional genotypes which have discovered some competitive advantage—and they are as unimportant to most statistical measures of diversity as they are fundamental to the process of evolution.

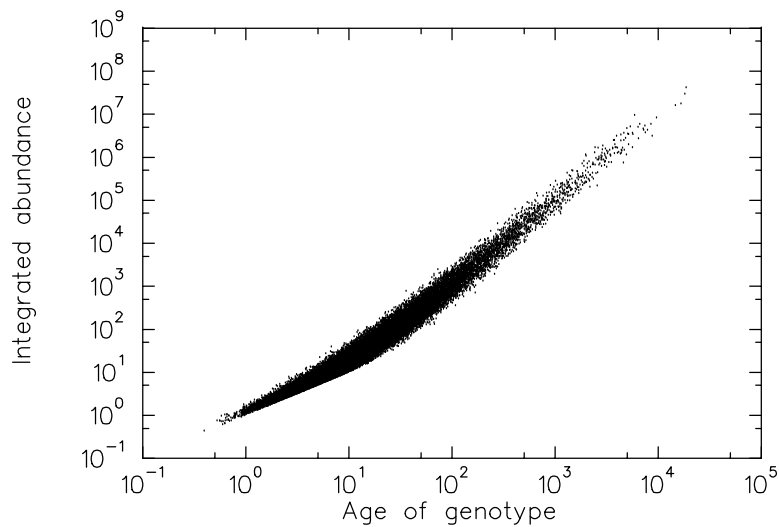

Fig. 7. Genotype age *vs.* integrated abundance for the DL model (analogous to Fig. 3).

## 4 Conclusions

Measuring abundance distributions in any living system is notoriously difficult, as witnessed by the limited amount of data in the literature, and reflected in a continuing uncertainty as to the nature of the distributions. The advent of Artificial Life systems analogous to simple natural systems has introduced the ability to study the old unanswered questions but with new methods and tools.

With Artificial Life it is now possible to gather huge amounts of population data from simple artificial evolving systems. From a simulation of an artificial world, not only the current ecology, but also the complete "paleontological" history can be recorded and examined in full detail. Parameters such as mutation rates and carrying capacities can also be adjusted easily.

In this paper we have re-examined some of the old questions by measuring both abundance distributions and the area-age relation in an Artificial Life system. Although it is currently not possible to distinguish species in these RNA-type systems, we are able to produce genotype abundance distributions showing universal behavior, and to speculate that the genotype distributions are the very foundation of the ubiquitous (-3/2) power-laws in species abundance curves obtained already in 1922 by Willis, and later throughout the taxonomic system and the paleontological record by Burlando.

Finally, we used our experience with the avida system to develop the DL stochastic model of population dynamics. This surprisingly simple model explains most of the dynamics of equilibrium periods, which in turn occupy the vast majority of the time of the simulations. It also provides accurate quantitative predictions for effects seen both in avida and in nature. In the future, we expect to refine the DL model to more accurately incorporate the spatial and finite-size effects present in the avida system, in order to isolate the unique features of an adaptive, evolving system.

### Acknowledgements

We would like to thank C. Ofria for discussions and collaboration in the design of avida. This research was supported in part by NSF grant PHY90-13248. C.A. acknowledges a Caltech Division Fellowship.